# Critical factor for epitaxial growth of cobalt-doped BaFe$_2$As$_2$ films by pulsed laser deposition

Hidenori Hiramatsu,[1,2,*] Hikaru Sato,[1] Takayoshi Katase,[3,**] Toshio Kamiya,[1,2] and Hideo Hosono[1,2,3]

[1] Materials and Structures Laboratory, Tokyo Institute of Technology, Mailbox R3-1, 4259 Nagatsuta-cho, Midori-ku, Yokohama 226-8503, Japan

[2] Materials Research Center for Element Strategy, Tokyo Institute of Technology, Mailbox S2-16, 4259 Nagatsuta-cho, Midori-ku, Yokohama 226-8503, Japan

[3] Frontier Research Center, Tokyo Institute of Technology, Mailbox S2-13, 4259 Nagatsuta-cho, Midori-ku, Yokohama 226-8503, Japan

PACS numbers: 81.15.Fg (laser ablation deposition), 74.70.Xa (Superconducting materials: pnictides), 74.78.-w (Superconducting films), 42.55.Lt (Excimer lasers), 74.25.F- (Superconductors: transport properties)

**Abstract**

We heteroepitaxially grew cobalt-doped $BaFe_2As_2$ films on $(La,Sr)(Al,Ta)O_3$ single-crystal substrates by pulsed laser deposition using four different wavelengths and investigated how the excitation wavelength and pulse energy affected growth. Using the tilting and twisting angles of X-ray diffraction rocking curves, we quantitatively analyzed the crystallinity of each film. We found that the optimal deposition rate, which could be tuned by pulse energy, was independent of laser wavelength. The high-quality film grown at the optimal pulse energy (i.e., the optimum deposition rate) exhibited high critical current density over 1 MA/cm$^2$ irrespective of the laser wavelength.

Footnote

[*] : Corresponding author. E-mail: h-hirama@lucid.msl.titech.ac.jp

[**] : Present address: Research Institute for Electronic Science, Hokkaido University, Sapporo 001-0020, Japan

The report of an iron-based superconductor in 2008[1] soon provoked extensive research on growing thin films of the related materials.[2–7] These materials are advantageous for high magnetic-field applications, such as superconducting wires and tapes, because they have high upper critical fields of >50 T,[8] small anisotropy factors,[9] and good grain boundaries.[10] Many researchers have reported superconducting wires[11–13] and tapes[14–16] made from these materials that exhibit high critical current densities ($J_c$) under high magnetic fields.

Among the iron-based superconductors, 122-type cobalt-doped $BaFe_2As_2$ [$Ba(Fe,Co)_2As_2$] has been extensively studied because of its chemical stability[17] and its ease of epitaxial growth compared with other iron-based compounds such as $LaFeAs(O,F)$ and $(Ba,K)Fe_2As_2$. These advantages originate from the lower vapor pressure of the Co dopant compared with those of F and K dopants. Researchers have been able to achieve high $J_c$ ($\geq 1$ MA/cm$^2$) in high-quality $Ba(Fe,Co)_2As_2$ epitaxial films,[18–20] leading to demonstrations of Josephson junctions[10,21,22] and superconducting quantum-interference devices[23] built from those films.

So far, $Ba(Fe,Co)_2As_2$ epitaxial films with high $J_c$ have been effectively grown on two kinds of buffer layers: perovskite-type oxides,[18] such as $SrTiO_3$, and metallic Fe.[24] These buffer layers relax the in-plane lattice mismatch between the single-crystal substrate and the $Ba(Fe,Co)_2As_2$ film. In contrast, we have grown these films with high $J_c$ on single-crystal substrates without buffer layers by optimizing the growth conditions of pulsed laser deposition (PLD).[19,25] Other researchers[18,24] employing buffer layers have used KrF excimer lasers as their PLD excitation source; in contrast, we used a neodymium-doped yttrium aluminum garnet (Nd:YAG) laser to produce high-performance $Ba(Fe,Co)_2As_2$ epitaxial films ever since we epitaxially grew the iron-based superconductor LaFeAsO.[2] The reason why the laser choice mattered is not yet clear because PLD has many parameters, such as the geometrical configuration, the base pressure of the growth chamber, the quality of the targets, and the excitation laser source. Among these parameters, we thought the important difference in our setup was the wavelength of the excitation laser used for PLD.

In this study, we grew Ba(Fe,Co)$_2$As$_2$ epitaxial films by using four types of ns-pulsed lasers; by growing films over a variety of pulse energies, we found the film crystallinity depended on the growth rate (pulse energy), rather than the excitation laser wavelength.

Ba(Fe,Co)$_2$As$_2$ films were deposited on (001)-oriented (La,Sr)(Al,Ta)O$_3$ (LSAT) single-crystal substrates without buffer layers. We deposited these films by using PLD to ablate Ba(Fe$_{0.92}$Co$_{0.08}$)$_2$As$_2$ target disks.[19,25] We used the same deposition chamber for every growth. We used four excitation sources for the pulsed laser: (i) an ArF excimer laser (wavelength $\lambda$ = 193 nm), (ii) a KrF excimer laser (248 nm), (iii) the second harmonic of a Nd:YAG laser (532 nm), and (iv) the fundamental harmonic of a Nd:YAG (1064 nm) laser. The COMPex 205 series (Lambda Physik, maximum pulse energies are 400 mJ for ArF and 700 mJ for KrF) was used for the excimer lasers, and the INDI-40 series (Spectra Physics, maximum pulse energies are 200 mJ for 532 nm and 450 mJ for 1064 nm) was used for the Q-switched Nd:YAG laser. The laser spots at the target surface were 1.5×2.0 mm rectangles for the excimer lasers and 2-mm-diameter circles for the Nd:YAG laser. The growth temperature was 850 °C.[25] In this study, we only varied the pulse energy. The pulse width and pulse energy were measured with a photodiode and an energy meter, respectively, which were calibrated for each wavelength. The repetition rate of each laser was 10 Hz. The distance between the substrate and the target was 30 mm. The base pressure of the PLD growth chamber was

~$5\times10^{-7}$ Pa. The film thicknesses were 200–300 nm, measured with a stylus surface profiler; while thickness of the thinner films (~90 nm) was determined by using X-ray reflectivity measurements.

Using out-of-plane and in-plane X-ray diffraction (XRD), we confirmed that all the films grew heteroepitaxially on the LSAT (001) substrates.[25] Variations of the crystallite orientation were characterized by XRD rocking curves of the out-of-plane 004 diffraction ($2\theta$–fixed $\omega$ scans, $\Delta\omega$ = tilting angle) and in-plane 200 diffraction ($2\theta_\chi$–fixed $\phi$ scans, $\Delta\phi$ = twisting angle). These measurements used Cu K$\alpha_1$ radiation with a Ge (220) monochromator. The samples' microstructures were observed by cross-sectional transmission electron microscopy (TEM). The chemical compositions of the films were analyzed with an energy dispersive X-ray (EDX) spectrometer with a spatial resolution of ~1 nm, attached to a scanning TEM.

The magnetic $J_c$ at 2 K up to 9 T was extracted using the Bean model from magnetization hysteresis loops, measured with a vibrating sample magnetometer. In these measurements, an external magnetic field ($H$) was applied normal to the substrate plane (i.e., parallel to the $c$-axis of the films). Optical transmission ($T_{obs}$) and normal reflectance ($R_{obs}$) spectra were measured with a conventional spectrophotometer at room temperature in the ultraviolet to near-infrared region. The absorption coefficient ($\alpha$) was evaluated from $T_{obs}$ and $R_{obs}$ by the following relationship: $T_{obs} / (1-R_{obs}) \approx \exp(-\alpha d)$, where $d$ is the film thickness (90 nm).

Fig. 1(a) shows the optical spectra of a 90-nm-thick Ba(Fe,Co)$_2$As$_2$ epitaxial film. In the wavelength region of Nd:YAG, the $\alpha$ values were $4.3\times10^5$ cm$^{-1}$ at $\lambda$ = 532 nm and $3.0\times10^5$ cm$^{-1}$ at 1064 nm, indicating that the pulse energy was absorbed down to several tens of nanometers below the surface (i.e., the penetration depth). These spectra are explained well by intraband transitions in the metallic band structure of Ba(Fe,Co)$_2$As$_2$.[26] In contrast, at the wavelength of the excimer lasers, the $\alpha$ values were > $1\times10^6$ cm$^{-1}$ and the penetration depths were a few nanometers.

Fig. 1(b) shows the deposition rate (DR) of the Ba(Fe,Co)$_2$As$_2$ film as a function of laser pulse energy. For Nd:YAG ablation at 1064 nm, the DR was extremely high compared with the other cases and grew linearly with pulse energy, but a distinct kink was present at a DR of ~3 Å/s. For Nd:YAG ablation at 532 nm, the DR was linear in the high-energy region, but a distinct kink was also present at ~3 Å/s (see the inset of Fig. 1(b) for a magnified view). We found similar trends also for ablation using the KrF excimer laser. Although the ArF excimer laser provided a maximum output pulse energy of 400 mJ, the maximum pulse energy irradiating the target was reduced to 165 mJ, mostly because of absorption of the deep ultraviolet light by O$_2$ molecules in the ambient air. In addition, for this laser, it was hard to observe a visible plume from the surface of the target disk at pulse energies less than 100 mJ, where the resulting thin films were very inhomogeneous and too thin. Therefore, for the ArF laser we measured the DR between 115 and 165 mJ. A similar phenomenon occurred also for the KrF ablation, where the minimum pulse energy required to form a plume was 50–60 mJ. These results indicate that the ablation threshold energy is higher for shorter wavelengths. Because Ba(Fe,Co)$_2$As$_2$ has a metallic band structure,[26] ablation should be dominated by the thermal effect, rather than electronic excitation such as multi-photon processes.[27] We believe that the high ablation threshold energy and low DR when using excimer laser excitation was caused by the very thin absorption layer.

Next, we examined how the crystallinity of the Ba(Fe,Co)$_2$As$_2$ film depended on pulse energy. Because all the films exhibited $c$-axis orientation in the out-of-plane XRD measurements and because they did not differ significantly in concentrations of impurities (e.g., Fe),[25] we examined in-plane $\phi$-scans of the 200 diffraction to confirm the heteroepitaxy. Because Ba(Fe,Co)$_2$As$_2$ has a tetragonal lattice, we expect a four-fold symmetry at 90° in this scan if the film lacks a rotational domain. However, for ArF ablation ((i) in Fig. 2(a)), we observed two kinds of domains rotated by 45° in all pulse-energy regions, indicating that ablation using the ArF excimer laser did not produce a high-quality epitaxial film. Also, for KrF ablation, we observed a similar rotational domain at a relatively low pulse energy ((ii) in Fig. 2(a)). This result is similar to that without a buffer layer reported by Lee et al.[18] employing the KrF excimer laser. However, we found that, by further increasing the pulse energy, Ba(Fe,Co)$_2$As$_2$ films exhibiting four-fold symmetry grew directly on the LSAT substrates, as shown in (iii) and (iv) in Fig. 2(a). For Nd:YAG ablation (Fig. 2(b)), we did not observe any rotational domain over the entire pulse-energy region we examined. We also evaluated the full width at half maximums (FWHMs) of the rocking curves for the out-of-plane (Fig. 2(c)) and in-plane (Fig. 2(d)) measurements as functions of pulse energy. These data have inverted bell-like shapes, and the minimum FWHMs fall in a range of $\Delta\omega$, $\Delta\phi$ = 0.6–0.7°; the optimum pulse energy depended strongly on the wavelength, shifting to higher energy with decreasing wavelength.

From the above results, we discuss why the crystallinity takes the optimum values against pulse energy. We believe that the DR of this process was dominated by the density of deposition precursors adsorbed on the growing surface and their re-evaporation rate; we can reasonably assume that the re-evaporation rate was constant for the range of laser power we used. Thus, the low DR at pulse energies lower than the kinks in Fig. 1(b) can be explained by the re-evaporation rate being comparable to the adsorption rate in that regime. In fact, the DRs at low pulse energy were smaller than expected from extrapolating the linear relationship from the higher pulse energies, as seen in the inset of Fig. 1(b), supporting the dependence of DR on re-evaporation in the low-energy regime. In other words, we believe the kinks observed at a DR of ~3 Å/s correspond to a transition to the supersaturation regime.[30] In contrast, at high pulse energy, the DR was too high to complete the surface reconstruction, leading to the increased FWHMs with increasing pulse energy, completing the inverted bell-like shape of the FWHM. Iida et al.[28,29] also reported epitaxial growth of Ba(Fe,Co)$_2$As$_2$ films using a KrF excimer laser, but they need to use a Fe buffer layer to obtain good epitaxial films. This would be because that their laser power was in the range of 3–5 J/cm$^2$, which is much lower than our optimum values (see Table I), and their substrate-to-target distance was longer (50 mm) than that of our PLD growth chamber (30 mm); therefore, we speculate that the above-discussed supersaturation regime is not attained due to the low density and the low kinetic energies of the deposition precursors, and consequently the Fe buffer layer is required to assist improved epitaxial growth.

Based on these results, the optimum range of pulse energy is 200–300 mJ (excitation density = 6.7–10 J/cm$^2$) for KrF, 70–100 mJ (2.2–3.2 J/cm$^2$) for Nd:YAG at 532 nm, and 40–50 mJ (1.3–1.6 J/cm$^2$) for Nd:YAG at 1064 nm.

Fig. 3(a) shows a cross-sectional TEM image of the Ba(Fe,Co)$_2$As$_2$ epitaxial film, deposited using the KrF excimer laser at an optimum pulse energy. Similar to films grown by Nd:YAG ablation at 532 nm,[31,32] we observed line defects along the *c*-axis, indicated by vertical white arrows, that act as vortex pinning centers. At the interface between the substrate and film, we observed a bright region with a thickness of a few nanometers, a feature more easily seen in Fig. 3(b). We found no differences from the KrF-deposited film in the defect structure or interface structure/contrast in the cross-sectional TEM images of the films deposited with the Nd:YAG laser at 532 and 1064 nm at optimal pulse energies (see Figs. S1 (a) and (b) in supplementary material for TEM images and EDX spectra[33]). Next, we show the interfacial chemical composition of the film deposited using the KrF excimer laser (Fig. 3(c)). The EDX intensities of Ba, Fe, Co, and As were almost constant in the deep film region; while they gradually decreased as the probing beam approached the interface, and elements from the substrate (La and Sr) were detected at the interface. The transition width was ~8 nm, much larger than the spatial resolution of this EDX measurement (~1 nm), which we attribute to diffusion. However, we found no segregation of specific elements, such as Fe, at the interface. This result differs from that reported by Iida *et al.*;[29] they observed a biaxially textured thin Fe layer at the interface and claimed that the thin Fe buffer layer was important to heteroepitaxially growing their Ba(Fe,Co)$_2$As$_2$ films.[24,34] Our present result is more similar to that reported by Rall *et al.*;[20] they reported a 2-nm-thick Fe-rich and Ba-poor reaction layer. We observed similar reaction layers also in the 532 and 1064 nm laser ablations. In previous work on MgO substrates,[36] we did not observe a thin reacted interfacial layer. Thus, we believe this interfacial phenomenon will commonly occur when epitaxially growing Ba(Fe,Co)$_2$As$_2$ films on LSAT substrates.

As seen in Fig. 2(a), our Ba(Fe,Co)$_2$As$_2$ films possess rotational domain structures when the DRs were low. Also in these cases, we always observed reaction layers similarly to those observed in the optimal samples (see TEM images in Figs. S1 (c) in supplementary material[33] and Fig. 3 for TEM images and EDX spectrum), but we could not find clear difference in their structures and compositions. However, atomic structures of the growing surfaces of these reaction layers should play an important role for determining the epitaxial structures of the growing thin films; *e.g.*, it is reported that *c*-plane α-Ga$_2$O$_3$ on *c*-plane α-Al$_2$O$_3$ [35] exhibit similar rotational domains because α-Al$_2$O$_3$ (0001) surfaces have different atomic structures that are rotated by 180 degrees with each other and the rotational domains are formed where a single-molecular layer step is formed at the substrate surface.

Fig. S2 in supplementary material[33] shows the magnetic $J_c$ at 2 K of the Ba(Fe,Co)$_2$As$_2$ epitaxial films deposited with KrF and Nd:YAG lasers at optimum pulse energies. Irrespective of the excitation wavelength used during growth, the films exhibited high self-field $J_c$ (>1 MA/cm$^2$), comparable to those of films grown on SrTiO$_3$ buffer layers[18] and higher than those of films grown on metallic Fe buffer layers.[37] The in-field properties of the film grown at 532 nm appeared slightly better than the others, implying a slightly higher defect density (i.e., density of pinning centers); however, the samples had similar decay ratios under magnetic fields. Thus, we conclude that high-$J_c$ Ba(Fe,Co)$_2$As$_2$ epitaxial films can be fabricated by using a variety of excitation wavelengths, assuming they are deposited at an optimum pulse energy.

Next, we discuss how the laser excitation parameters affected crystallinity of Ba(Fe,Co)$_2$As$_2$ films. Table I summarizes the optimum pulse energy for each laser to fabricate Ba(Fe,Co)$_2$As$_2$ epitaxial films with high $J_c$ as well as related optical parameters. Although the optimal pulse energy and pulse width for each laser were very different, the DRs produced by those optimum conditions were almost the same (3.3 ± 0.5 Å/s), indicating that the most important growth parameter was the DR. This finding explains why a high-$J_c$ Ba(Fe,Co)$_2$As$_2$ film with the ArF excimer laser could not be produced: its maximum pulse energy was limited to 165 mJ (5.5 J/cm$^2$), producing a maximum DR (2.3 Å/s) far lower than the optimum value (3.3 Å/s). As explained before, the optimum pulse energies and thus the excitation energy densities differed significantly between the lasers; in contrast, the photon number PN ($10^{17}$) and peak power density PPD ($10^8$ W/cm$^2$) were on the same order of magnitude for all the wavelengths used. To further discuss these parameters, we re-plot the data from Figs. 2(c) and (d) in Fig. S3 (See supplemental material for these re-plots[33]) with respect to $(1-R_{obs})$PN and $(1-R_{obs})$PPD, where the PN and PPD are corrected with the actual photon count absorbed by the Ba(Fe,Co)$_2$As$_2$ ($R_{obs}$ values are taken from Fig. 1(a)). As for $(1-R_{obs})$PN (Figs. S3(a) and (b)), the optimum range is narrow, (1.3–3)×10$^{17}$, while the values for ablation at 248 nm are higher than those for ablation at longer wavelengths. As for $(1-R_{obs})$PPD (Figs. S3(c) and (d)), the optimum range is again narrow, but the value for ablation at 1064 nm deviates much from the others. These results suggest that $(1-R_{obs})$PN is closely correlated with the DR. However, although this result implies that the DR (i.e., the ablation rate) was determined by a single-electron excitation process, this explanation is not consistent with previous research on the mechanisms of laser ablation. In the previous research, electron excitation processes become dominant at shorter pulse widths (larger energy densities) and are important for insulator and semiconductor films. In contrast, when ablating metallic materials like Ba(Fe,Co)$_2$As$_2$ with nanosecond pulses, the thermal process should dominate.[27] Therefore, we tentatively believe that the DR is mostly determined by the absorbed PPD. The deviation observed for ablation at 1064 nm can be explained by its large penetration depth (33 nm) compared with ablation at 532 nm (23 nm) and 248 nm (~5 nm). The 1064 nm laser ablated a larger amount of the PLD target, resulting in a high ablation efficiency and very high DR, as shown by Fig. 1(b); in contrast, the lasers with shorter wavelengths ablated thinner surfaces, resulting in lower DRs even at the same PPD.

In summary, we epitaxially grew Ba(Fe,Co)$_2$As$_2$ films by using PLD and examined how the laser wavelength and pulse energy affected the growth by using four different excitation wavelengths. We found that the optimal DR, which could be tuned by pulse energy, does not depend on the type of laser (i.e., wavelength). This study also explains why the Nd:YAG laser is better for producing high-$J_c$ Ba(Fe,Co)$_2$As$_2$ films with high crystallinity at a low laser power, and will help improve the fabrication of other iron-based superconductor films such as *RE*FeAsO (*RE* = rare earth), BaFe$_2$(As,P)$_2$, and Fe(Se,Te).

**Acknowledgments**

This work was supported by the Japan Society for the Promotion of Science (JSPS), Japan, through the "Funding Program for World-Leading Innovative R&D on Science and Technology (FIRST Program)", and the Ministry of Education, Culture, Sports, Science and Technology (MEXT) Element Strategy Initiative to Form Core Research Center. H. Hiramatsu was also supported by a JSPS Grant-in-Aid for Young Scientists (A) Grant Number 25709058 and a JSPS Grant-in-Aid for Scientific Research on Innovative Areas "Nano Informatics" Grant Number 25106007.

Table I. Optimum pulse energies and deposition rates for each laser to fabricate Ba(Fe,Co)$_2$As$_2$ epitaxial films with high $J_c$ as well as related optical parameters.

| Laser wavelength (nm) | 248 | 532 | 1064 |
| --- | --- | --- | --- |
| $R_{obs}$ (%) [a] | 24 | 34 | 52 |
| Pulse width (ns) | 20 | 5 | 10 |
| Spot area ($10^{-2}$ cm$^2$) | 3.0 | 3.1 | 3.1 |
| Optimum pulse energy (mJ) | 200 – 300 | 70 – 100 | 40 – 50 |
| Deposition rate (Å/s) | 3.1 – 3.6 | 2.8 – 3.3 | 3.5 – 3.8 |
| Photon number per pulse, PN ($10^{17}$) [b] | 2.5 – 3.8 | 1.9 – 2.7 | 2.2 – 2.7 |
| Excitation energy density (J/cm$^2$) [c] | 6.7 – 10 | 2.2 – 3.2 | 1.3 – 1.6 |
| Peak power density, PPD ($10^8$ W/cm$^2$) [d] | 3.3 – 5.0 | 4.5 – 6.4 | 1.3 – 1.6 |

a: taken from FIG. 1(a), b: Photon number per pulse = Pulse energy (J) / Photon energy (J), c: Excitation energy density (J/cm$^2$) = Pulse energy (J) / Spot area (cm$^2$), d: Peak power density (W/cm$^2$) = Pulse energy (J) / Pulse width (s) / Spot area (cm$^2$).

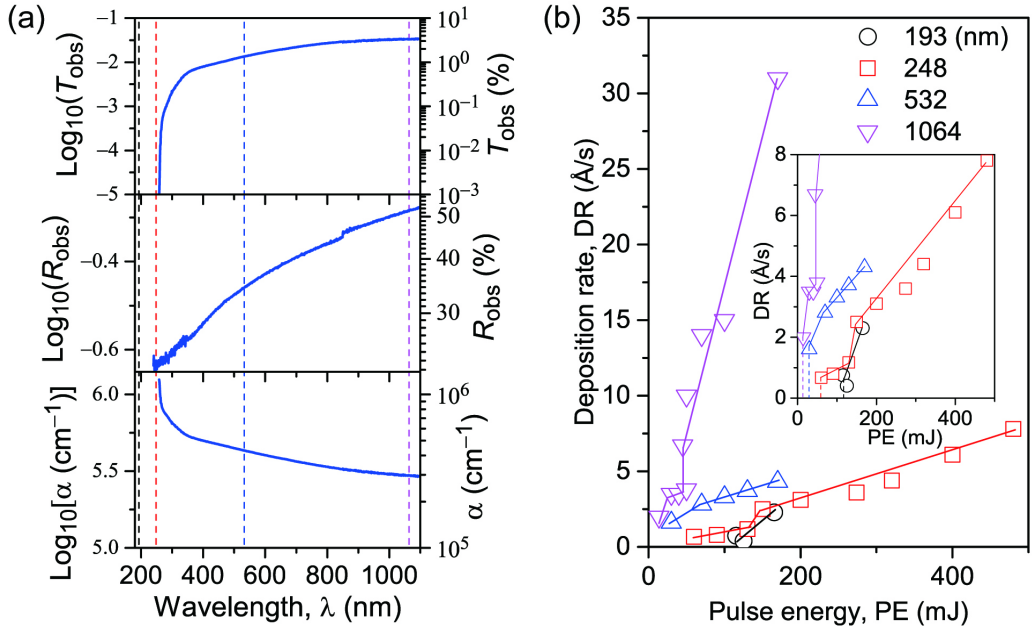

FIG. 1. (Color online) (a) Optical spectra ($T_{obs}$, $R_{obs}$, and $\alpha$) of a 90-nm-thick Ba(Fe,Co)$_2$As$_2$ epitaxial film at room temperature. The vertical dashed lines indicate the laser wavelengths used in this study. (b) Dependence of deposition rate of Ba(Fe,Co)$_2$As$_2$ films on pulse energy. The PLD laser wavelengths are shown in the upper right. The inset shows an enlarged image from the region of deposition rates less than 8 Å/s, more clearly showing the kinks.

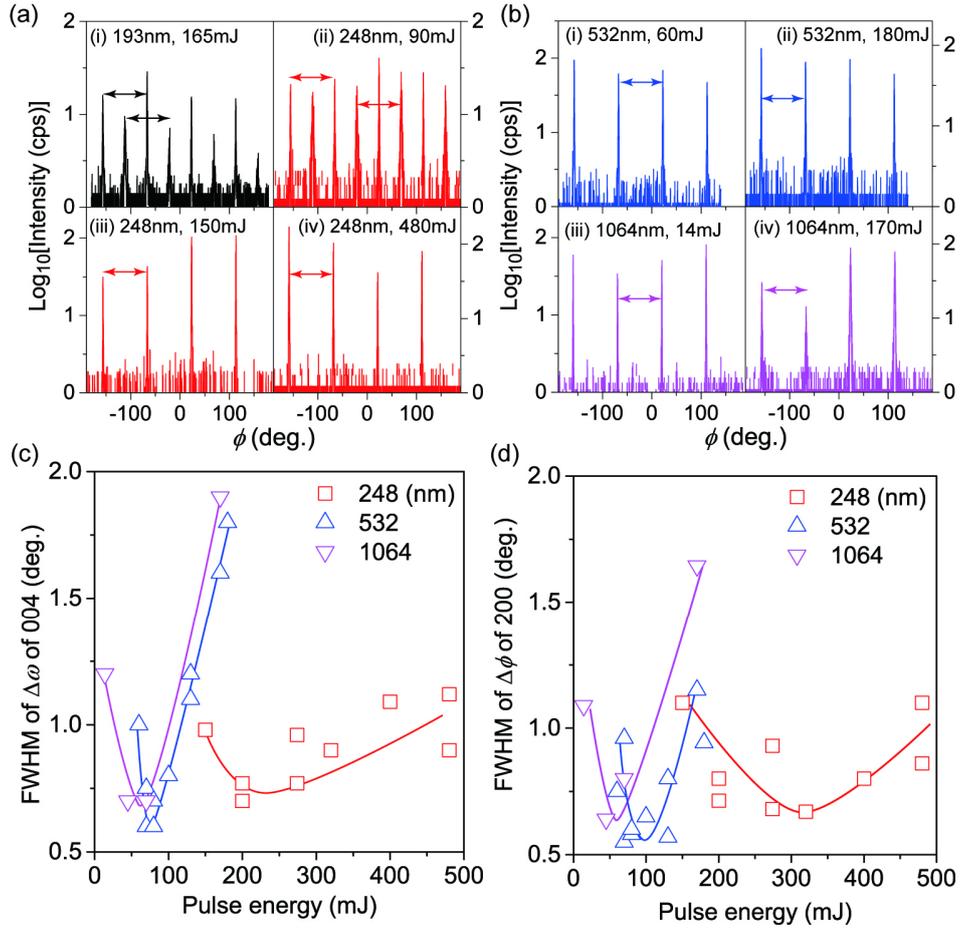

FIG. 2. (Color online) Relationship between thin film crystallinity and pulse laser energy. (a) and (b): XRD results from in-plane $\phi$-scans of 200 diffraction of Ba(Fe,Co)$_2$As$_2$ films grown by (a) excimer and (b) Nd:YAG lasers. The horizontal arrows show a 90° interval because of the film's tetragonal symmetry. (c) and (d): FWHMs of rocking curves of (c) out-of-plane 004 diffraction and (d) in-plane 200 diffraction of Ba(Fe,Co)$_2$As$_2$ epitaxial films without a rotational domain, grown by KrF and Nd:YAG lasers as a function of various pulse energies.

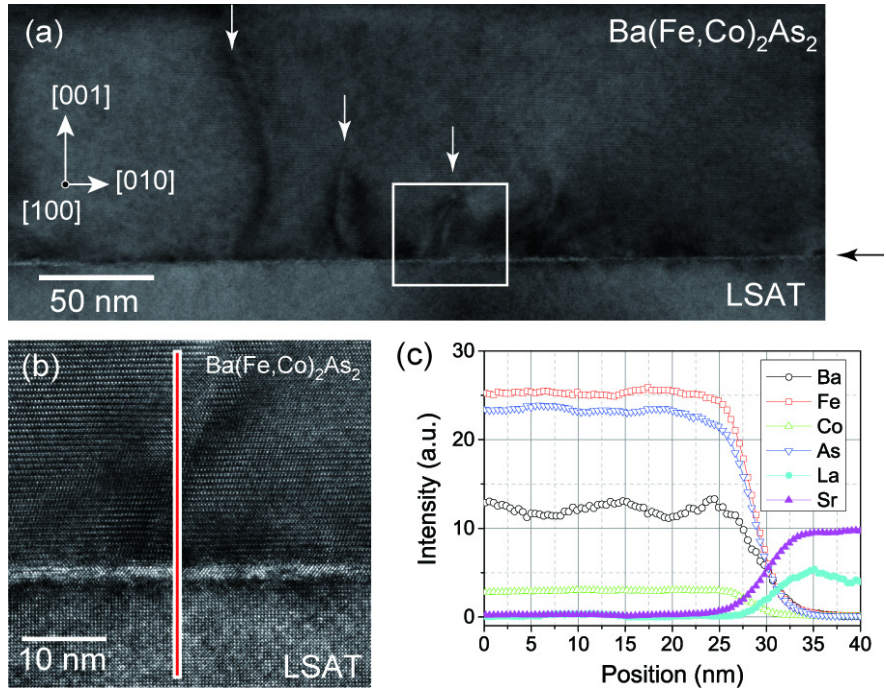

FIG. 3. (Color online) Examination of a Ba(Fe,Co)$_2$As$_2$ epitaxial film grown by the KrF excimer laser at the optimum pulse energy. (a) Cross-sectional TEM image. The vertical white arrows indicate defects along the *c*-axis. The horizontal black arrow on the right shows the position of the film/LSAT interface. (b) Magnified view of the white square shown in (a). (c) EDX line-scan spectra along the red vertical line shown in (b).

Supplementary material for "Critical factor for epitaxial growth of cobalt-doped BaFe$_2$As$_2$ films by pulsed laser deposition"

Hidenori Hiramatsu,[1,2] Hikaru Sato,[1] Takayoshi Katase,[3] Toshio Kamiya,[1,2] and Hideo Hosono[1,2,3]

[1] Materials and Structures Laboratory, Tokyo Institute of Technology, Mailbox R3-1, 4259 Nagatsuta-cho, Midori-ku, Yokohama 226-8503, Japan

[2] Materials Research Center for Element Strategy, Tokyo Institute of Technology, Mailbox S2-16, 4259 Nagatsuta-cho, Midori-ku, Yokohama 226-8503, Japan

[3] Frontier Research Center, Tokyo Institute of Technology, Mailbox S2-13, 4259 Nagatsuta-cho, Midori-ku, Yokohama 226-8503, Japan

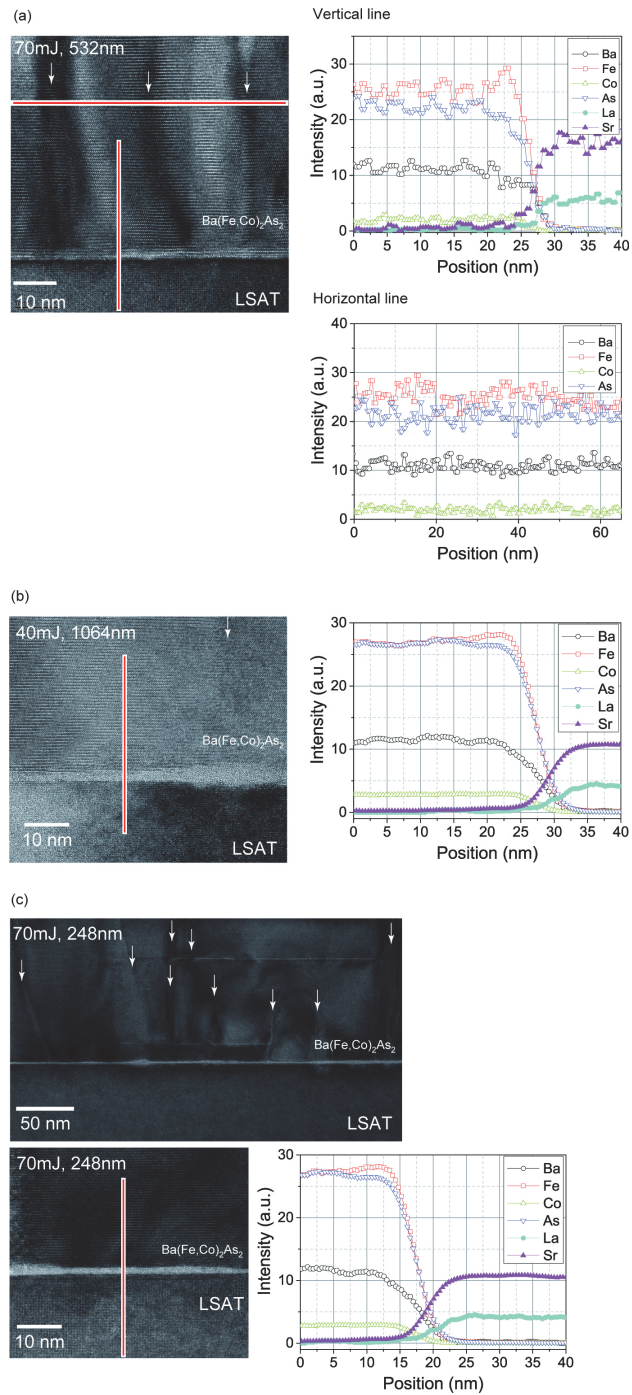

Supplementary FIG. S1. (left) Cross-sectional TEM images of Ba(Fe,Co)$_2$As$_2$ films grown by ablations at (a) optimum pulse energy (70 mJ) of 532 nm, (b) optimum pulse energy (40 mJ) of 1064 nm, and (c) low pulse energy (70 mJ) of 248 nm. The vertical white arrows indicate defects along the *c*-axis. (right) EDX line-scans along the red lines in the TEM images.

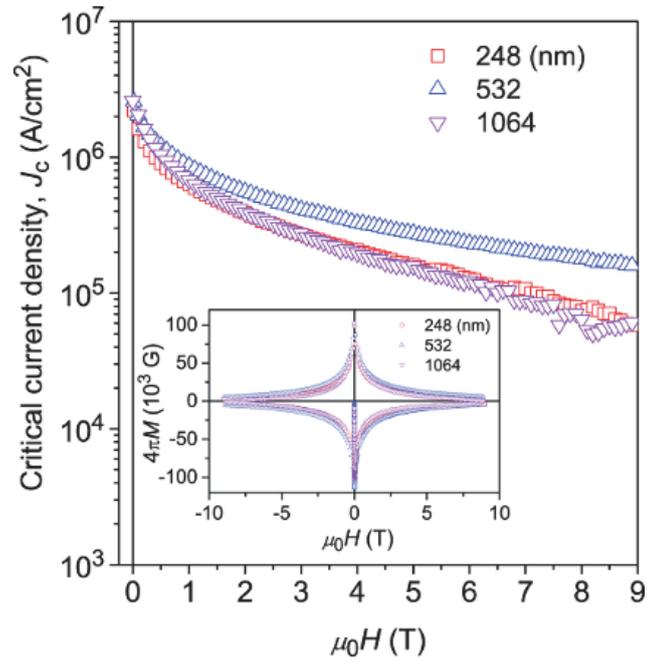

Supplementary FIG. S2. Magnetic $J_c$ at 2 K of Ba(Fe,Co)$_2$As$_2$ epitaxial films grown using three types of lasers, at their respective optimal pulse energies, as a function of magnetic field. The inset shows the magnetization ($M$) hysteresis loops measured at 2 K used to extract the magnetic $J_c$ for each sample.

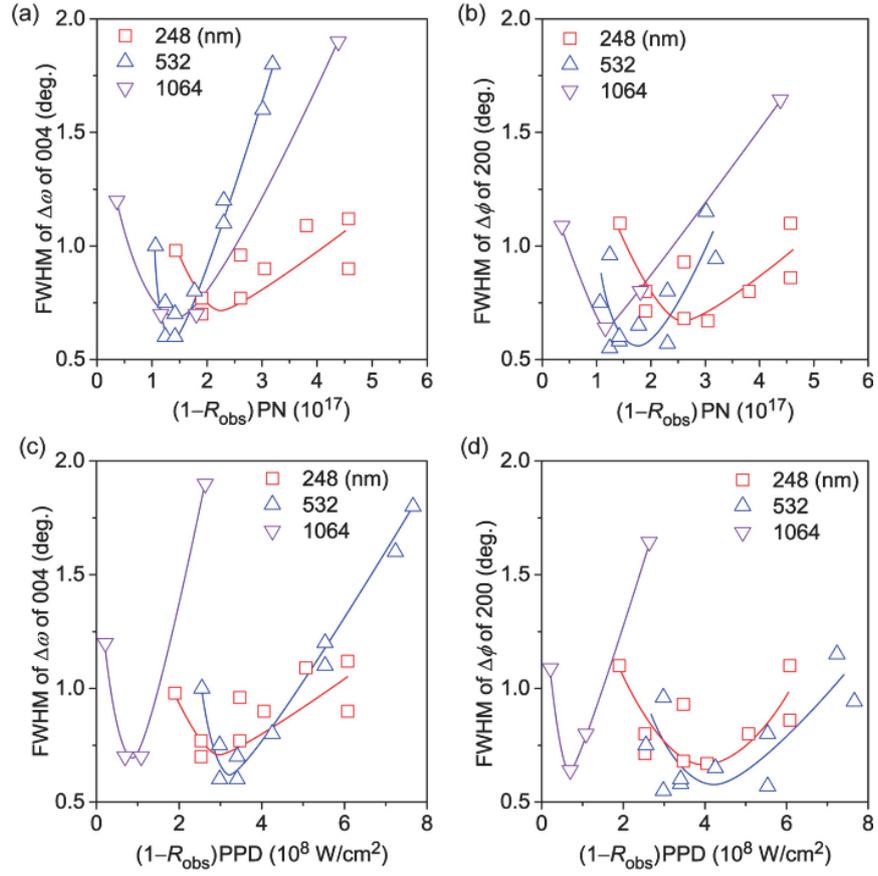

Supplementary FIG. S3. Re-plots of Figs. 2(c) and (d) with respect to $(1-R_{obs})$PN (a and b) and $(1-R_{obs})$PPD (c and d). Each $R_{obs}$ is taken from Fig. 1(a).